\documentclass[aps,prl,twocolumn,superscriptaddress,showpacs]{revtex4}

\usepackage{graphicx}

\begin{document}


\title{Electromagnetically Induced Transparency with an Ensemble of\\
Donor-Bound Electron Spins in a Semiconductor}


\author{Maksym~Sladkov}

\author{A.~U.~Chaubal}

\author{M.~P.~Bakker}

\author{A.~R.~Onur}

\affiliation{Zernike Institute for Advanced Materials,
University of Groningen, NL-9747AG Groningen, The
Netherlands}

\author{D.~Reuter}
\author{A.~D.~Wieck}
\affiliation{Angewandte Festk\"{o}rperphysik,
Ruhr-Universit\"{a}t Bochum, D-44780 Bochum, Germany}

\author{C.~H.~van~der~Wal}
\affiliation{Zernike Institute for Advanced Materials,
University of Groningen, NL-9747AG Groningen, The
Netherlands}


\date{\today}

\begin{abstract}
We present measurements of electromagnetically induced
transparency with an ensemble of donor-bound electrons in
low-doped $n$-GaAs. We used optical transitions from the
Zeeman-split electron spin states to a bound trion state in
samples with optical densities of 0.3 and 1.0. The electron
spin dephasing time $T_2^* \approx 2~{\rm ns}$ was limited
by hyperfine coupling to fluctuating nuclear spins. We also
observe signatures of dynamical nuclear polarization, but
find these effects to be much weaker than in experiments
that use electron spin resonance and related experiments
with quantum dots.
\end{abstract}


\maketitle


A localized electronic spin in a semiconductor is a
promising candidate for implementing quantum information
tasks in solid state. Optical manipulation of
single-electron and single-hole systems has been realized
with quantum dots
\cite{Press2008,Greilich2009,Xu2009,brunner2009science,Latta2009a}
and by using donor atoms that are not ionized at low
temperature ($D^0$ systems) \cite{Fu2005,Fu2008,Clark2009}.
These results illustrate the potential of quantum-optical
control schemes that come within reach when adapting
techniques from the field of atomic physics. An advantage
of the $D^0$ systems over dots is that these can be
operated as an ensemble with very little inhomogeneity for
the optical transition energies. Such ensembles at high
optical density are key for robust quantum-optical control
schemes that have been designed for preparing nonlocal
entanglement between spins, quantum communication, and
applying strong optical nonlinearities
\cite{Duan2001,Fleischhauer2005a}. A critical step toward
implementing these schemes is the realization of
electromagnetically induced transparency (EIT). We present
here measurements of EIT with an ensemble of donor-bound
electron spins in low-doped $n$-GaAs, in samples with
optical densities of 0.3 and 1.0 \cite{Wang2007}. We build
on an earlier indirect observation of coherent population
trapping with this system \cite{Fu2005}. Extending this to
a direct realization of EIT with an optically dense medium
is essential for getting access to strong field-matter
interactions without using optical cavities, and for the
application and study of transmitted signal fields
\cite{Duan2001,Fleischhauer2005a}.

\begin{figure}
  \includegraphics[width=86mm]{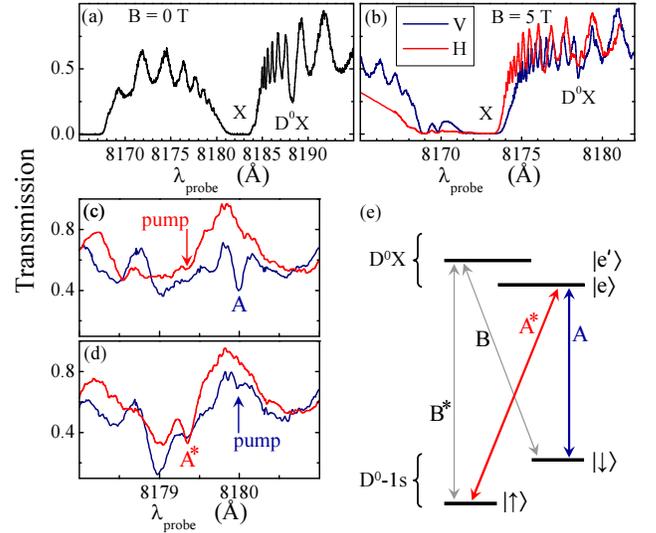}\\
  \caption{(color)(a) Transmission spectroscopy at $B=0$~T.
  (b) Transmission at $B=5.0$~T for H and V polarization.
  (c) Pump-assisted spectroscopy with H-polarized
  pumping at the $A^*$ transition shows enhanced
  absorption for the $A$ transition for the scan with a V-polarized
  probe (blue trace), but not with an H-polarized
  probe (red trace).
  (d) Complementary to (c), V-polarized pumping at $A$
  shows enhanced absorption for the $A^*$ transition with
  an H-polarized probe. (e)  Energy levels and optical transitions
  of the $D^0$-$D^0 X$ system.}\label{FullRangeSpectra}
\end{figure}

We implemented EIT in its most typical form where a spin-up
and a spin-down state ($\mid\uparrow\rangle$ and
$\mid\downarrow\rangle$ of the electron in the $D^0$
system) have an optical transition to the same excited
state $|e\rangle$ (Fig.~\ref{FullRangeSpectra}(e)). We
Zeeman-split the states $\mid\uparrow\rangle$ and
$\mid\downarrow\rangle$ with an applied magnetic field. For
the state $|e\rangle$ we used the lowest energy level of a
donor-bound trion system ($D^0 X$, with two electrons in a
singlet state and a spin-down hole with $m_{h}=-\frac12$
\cite{Clark2009} localized at the donor site). EIT is then
the phenomenon that absorbtion by one of the optical
transitions is suppressed because destructive quantum
interference with the other transition prohibits populating
the state $|e\rangle$. The $D^0$ systems are then trapped
in a dark state that is in the ideal case a coherent
superposition of the states $\mid\uparrow\rangle$ and
$\mid\downarrow\rangle$ only \cite{Fu2005,Wang2007}. This
state is proportional to $\Omega_{c} \mid\uparrow\rangle -
\Omega_{p} \mid\downarrow\rangle$, with $\Omega_{c}$ and
$\Omega_{p}$ the Rabi frequencies of the control and probe
field that drive the two transitions
\cite{Fleischhauer2005a}.

We present results of implementing EIT in GaAs, and we
studied the interactions between the solid-state
environment and driving EIT. In particular, the $D^0$
systems have a single electron in a hydrogen-like $1s$
wavefunction with a Bohr radius of $\sim10~{\rm nm}$, and
each electron spin has hyperfine coupling to $\sim 10^5$
fluctuating nuclear spins. We studied how this limits the
electron spin dephasing time and how driving EIT can result
in dynamical nuclear polarization (DNP). In addition, we
find that it is crucial to suppress heating effects from
the nearby free exciton resonance, and demonstrate that
with direct heat sinking of GaAs layers EIT can be driven
with $\Omega_{c}/2\pi$ up to 2~GHz, while keeping the spin
dephasing time $T_2^* \approx 2~{\rm ns}$ near the level
that results from the nuclear spin fluctuations.

\begin{figure}
  \includegraphics[width=86mm]{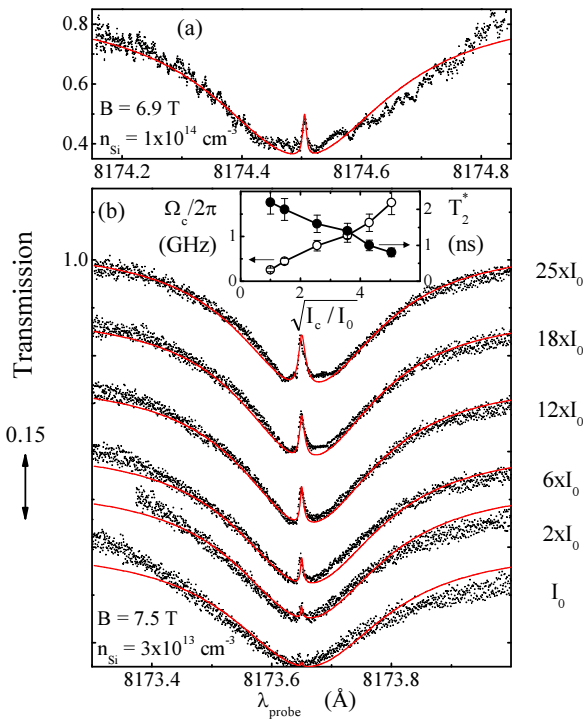}\\
  \caption{(color online)
   (a) EIT spectrum from sample with Si doping at
   $1 \times 10^{14}~{\rm cm^{-3}}$.
   Dots - experiment. Line - numerical fit.
   (b) EIT spectra from sample with Si doping
   at $3 \times 10^{13}~{\rm cm^{-3}}$, for
   probe-field intensity $0.04~{\rm W cm^{-2}}$ and
   a range of control-field intensities $I_{c}$
   with $I_0=0.4~{\rm W cm^{-2}}$. The inset shows
   the fitting results for Rabi
   frequency $\Omega_c$ and spin dephasing
   time $T_2^*$.}\label{PowerDep}
\end{figure}


We used epitaxially grown GaAs films of $10 \; {\rm \mu m}$
thickness with Si doping at $n_{\rm Si}=3\times10^{13}$ and
$1\times10^{14}~{\rm cm}^{-3}$. At these concentrations the
wavefunctions of neighboring donor sites do not overlap,
which yields an ensemble of non-interacting $D^0$ systems.
The films were transferred to a wedged sapphire substrate
with an epitaxial lift-off process \cite{Yablonovitch1987},
and fixed there by Van der Waals forces which assures high
heat sinking. The sapphire substrate was mounted on the
copper cold finger of a bath cryostat (4.2~K) in the center
of a superconducting magnet with fields $B$ up to 8~T in
the plane of the sample ($z$-direction). Laser light was
brought to the films at normal incidence (Voigt geometry)
via a polarization-maintaining single-mode fiber. The two
linear polarizations supported by the fiber are set
parallel (V polarization) and orthogonal (H polarization)
to the applied magnetic field. The V polarization can drive
$\pi$ transitions (no change of $z$-angular momentum) and
the H polarization can drive transitions with a change in
$z$-angular momentum of $\pm \hbar$.

Two CW Ti:sapphire lasers (Coherent MBR-110, linewidth
below 1~MHz) provided tunable probe and control fields.
Focussing in the sample volume was achieved with a
piezo-motor controlled confocal microscope. During
transmission experiments we defocussed the microscope to a
spot of $\sim$16~${\rm \mu m}$ diameter to avoid
interference effects from the cavity that is formed between
the sample surface and the facet of the fiber. The probe
field was amplitude modulated at $6~{\rm kHz}$ and we used
lock-in techniques for detecting light that is transmitted
trough the sample with a photodiode directly behind the
sample. The signal due to unmodulated control field is
rejected by AC coupling of the measurement electronics.


We first report transmission experiments that identify the
spectral position of the $D^0X$ related resonances. Only
the probe laser was used. Figure~\ref{FullRangeSpectra}(a)
shows a spectrum taken at $B=0~{\rm T}$ (identical result
for H and V polarization), and
Fig.~\ref{FullRangeSpectra}(b) shows a result for
$B=5.0~{\rm T}$ with a separate trace for H and V
polarization. The strong absorbtion labeled $X$ is due to
excitation of free excitons. Resonant absorption by
donor-bound excitons ($D^0X$) occurs at $8187.5~{\rm \AA}$
for $B=0~{\rm T}$ and at $8179.5~{\rm \AA}$ for $B=5.0~{\rm
T}$. The shift of the resonances with magnetic field is the
diamagnetic shift. The spacing of $5~{\rm \AA}$ between the
$X$ and $D^0X$ resonances is in good agreement with
previously reported binding energies
\cite{Bogardus1968,Karasyuk1994}. The oscillating
background superimposed on the resonances is due to a
Fabry-Perot effect in the GaAs film, and its chirped
wavelength dependence around $X$ is due to the wavelength
dependent refractive index that is associated with the
strong free exciton absorption.


For identifying the $A$ and $A^*$ transitions of
Fig.~\ref{FullRangeSpectra}(e) within the fine structure of
$D^0X$ spectra at high fields we performed scanning-probe
laser spectroscopy while the control laser is applied for
optical pumping of a particular $D^0X$ transition (this
also eliminates bleaching by the probe).
Figure~\ref{FullRangeSpectra}(c) shows spectra obtained
with pumping at $A^*$ ($8179.3~{\rm \AA}$) with H
polarization. This leads to enhanced absorbtion at the $A$
resonance $(8180.0~{\rm \AA})$ for the probe scan with V
polarization. The complementary experiment with pumping
V-polarized light into this $A$ transition leads to
enhanced absorption of H-polarized light at transition
$A^*$ (Fig.~\ref{FullRangeSpectra}(d)). We could also
perform such cross-pumping experiments using the $B$ and
$B^*$ transitions to the level $|e ' \rangle$ (the first
excited state of the series of energy levels of the $D^0X$
complex, see Fig.~\ref{FullRangeSpectra}(e)). We thus
confirmed that the pair of transitions labeled as $A$ and
$A^*$ address a so-called closed three-level
$\Lambda$-system, and that this is the pair with lowest
energies within the $D^0X$ resonances. This interpretation
is also consistent with the polarization dependence of
these transitions \cite{Karasyuk1994,Fu2005}. In the field
range 5 to 8~T, the $A$ and $A^*$ transitions are
spectrally well separated from the transitions $B$, $B^*$,
and transitions to higher excited states of the $D^0 X$
complex. The observed $D^0$ Zeeman splitting corresponds to
an electron $g$ factor $|g|=0.42$, and also agrees with
previous reports \cite{Karasyuk1994,Fu2005}.


We now turn to the observation of EIT
(Fig.~\ref{PowerDep}). For these results we fixed the
control laser central on the $A$ transition (V
polarization), while the probe laser is scanned across the
$A^*$ transition (H polarization). When the control and
probe field meet the condition for two-photon Raman
resonance (the difference in photon energy exactly matches
the $D^0$ spin splitting), a narrow peak with enhanced
transmission appears inside the broader $A^*$ absorption
dip, which is the fingerprint of EIT. In
Fig.~\ref{PowerDep}(a) this occurs inside an $A^*$
absorption with optical density 1.0, while for the sample
with $n_{\rm Si}=3\times10^{13}~{\rm cm}^{-3}$ this is 0.3
(Fig.~\ref{PowerDep}(b)). We further focus on this latter
sample since higher resolution of the EIT spectra makes it
more suited for our further studies.

\begin{figure}
  \includegraphics[width=86mm]{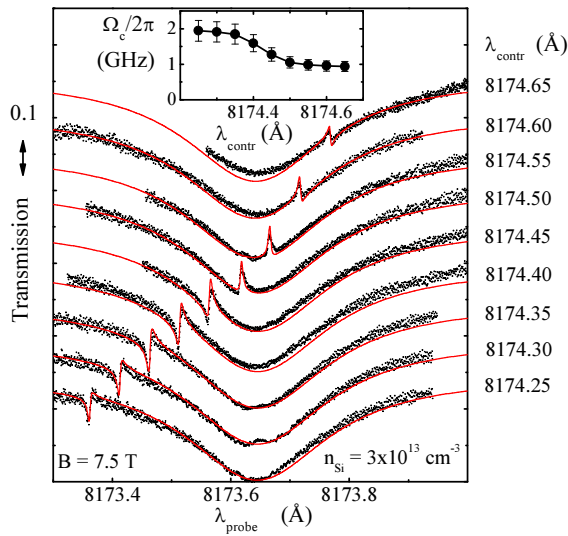}\\
  \caption{(color online) Dependence of EIT
  spectra on control-field detuning.
  The position of the EIT peak follows
  precisely the control-field detuning
  from transition $A$. Dots - experiment
  with control (probe) intensity $6~(0.04)~{\rm W cm^{-2}}$.
  Lines - fits with $T_2^*=2~\rm{ns}$ and $\Omega_c$ as
  presented in the inset.}\label{TwoPhotonDet}
\end{figure}



The lines in Fig.~\ref{PowerDep} and \ref{TwoPhotonDet} are
results of fitting EIT spectra with the established theory
\cite{Fleischhauer2005a}. This involves calculating the
steady-state solution of a density-matrix equation for the
three-level system, and accounts for coherent driving by
the lasers and relaxation and dephasing rates between the
levels. The free parameters are the inhomogeneous
broadening $\gamma_{A^*}$ (typically 6~GHz) for the optical
transition $A^*$, the spin dephasing time $T_2^*$ and the
control-field induced Rabi frequency $\Omega_c$ (and
$\Omega_p << \Omega_c$). The rest of the parameters are the
same as in Ref.~\cite{Fu2005}, and we found $\Omega_c$
always consistent with an independent estimate from the
optical intensity and electric dipole moment. We obtain
good fits and the main features in our results are
consistent with EIT, as we discuss next.


Figure~\ref{PowerDep}(b) shows EIT spectra taken at
different intensities $I_{c}$ of the control field, where a
stronger control field yields a higher and broader EIT
peak. As expected for EIT, we observe that $\Omega_c$ from
fits scales linearly with $\sqrt{I_{c}}$
(Fig.~\ref{PowerDep}(b), inset). The $\Omega_c$ values
reach $2\pi \cdot 2~\rm{GHz}$, and we could only obtain
clear EIT spectra with such high $\Omega_c$ in samples with
complete adhesion onto the sapphire substrate. Our results
from samples with incomplete adhesion (and work with
epi-layers that are not removed from the original GaAs
substrate \cite{Fu2005,Fu2008,Clark2009}) suffer from
heating, which is observed as a broadening of the free
exciton line into the region of the $D^0X$ resonances. The
values of $T_2^*$ that we find in our experiments are
discussed below.


Figure~\ref{TwoPhotonDet} shows how the EIT peak position
depends on detuning of the control field from the $A$
transition. As expected, the EIT peak follows the detuning
of the control field. However, the EIT peak in the
blue-detuned traces is clearly more prominent than in the
red-detuned cases. We attribute this to a change in the
effective Rabi frequency $\Omega_c$ that results from the
weak Fabry-Perot interference within the GaAs film, and we
can indeed fit the results with fixed $T_2^* = 2~{\rm ns}$
and varying $\Omega_c$ (Fig.~\ref{TwoPhotonDet}, inset). We
can exclude that the difference in the quality of EIT
spectra is coming from optical coupling to a level outside
our $\Lambda$-system, since all other transitions are well
separated spectrally and in polarization dependence
(\textit{e.g.} the $B$ and $B^*$ transitions, see
Fig.~\ref{FullRangeSpectra}(e)).


An important topic that needs to be addressed next with
this realization of EIT concerns the influence of the
hyperfine coupling between each electron spin and $\sim
10^5$ nuclear spins. A polarization of the nuclear spins
acts on the electron spin as an effective magnetic field
$B_{nuc}$. The average polarization affects the Zeeman
splitting, and this can be directly observed in EIT spectra
as a red (blue) shift of the EIT peak for a reduced
(enhanced) Zeeman splitting. The nuclear spin fluctuations
around the average dominate via this mechanism the
inhomogeneous electron spin coherence time $T_2^*$. This is
a key parameter for the shape of the EIT peak (longer
$T_2^*$ gives a sharper peak), and the magnitude of these
fluctuations can therefore be derived from the EIT spectra
as well. At our fields and temperature nuclear spins are in
equilibrium close to full random orientation. The expected
value for $T_2^*$ for this case is $\sim 2~{\rm ns}$
\cite{Fu2005,Kennedy2006}, and is in agreement with the
values that we observe.

The hyperfine coupling can also result in dynamical nuclear
polarization (DNP), which is the transfer of angular
momentum from the electron to the nuclear spins when the
electron spin is driven out of equilibrium. Earlier
experiments on our type of $D^0$ system with
microwave-driven electron spin resonance (ESR)
\cite{Kennedy2006} and optical experiments on quantum dots
showed strong DNP \cite{Xu2009,Latta2009a}. In both cases
the effects were so strong that it gave an unstable
resonance condition for tuning at ESR and EIT (the systems
trigger a DNP cycle that drives them out of resonance). DNP
can also result in a suppression of the nuclear spin
fluctuations, which yields a longer $T_2^{*}$
\cite{Greilich2009,Xu2009,Latta2009a,vink2009natphys}. Our
experiment, however, only shows weak DNP. We never observed
a significant change in the Zeeman energy (as derived from
subtracting the probe and control photon energies at the
EIT peak) from the EIT driving itself. We only observed in
several data sets a moderate EIT peak narrowing over the
course of a few hours of data taking (at fixed settings of
the EIT parameters). In order to confirm the role of
nuclear spins we carried out various attempts to induce
stronger DNP effects.

\begin{figure}
  \includegraphics[width=86mm]{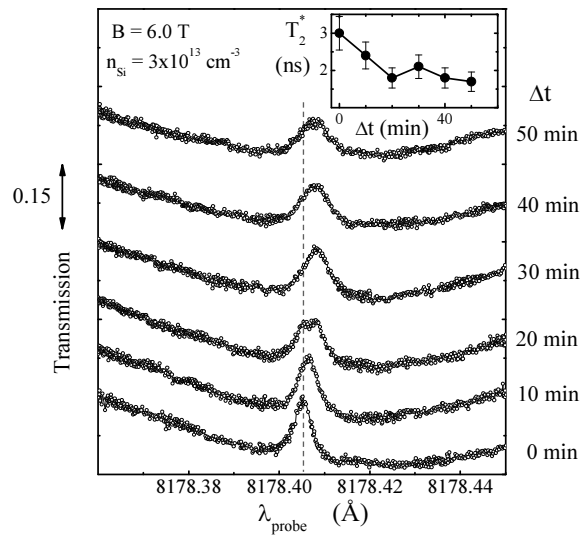}\\
  \caption{Evolution of the EIT peak after
  30~min pumping of the $A^*$ transition.
  Fast EIT 'snapshots' were taken at 10~min
  intervals during which the sample was kept
  in the dark. The dashed line is a guide for
  showing the shift in peak position. The inset
  presents fitting results that
  show the change in $T_2^*$. }
  \label{DNPpump}
\end{figure}

An example of the strongest DNP effects that we could
induce is presented in Fig.~\ref{DNPpump}. Here we first
applied strong driving of the $A^*$ transition for 30~min
with an intensity equivalent to a Rabi frequency of $2\pi
\cdot 10~\rm{GHz}$. This pumps the system fully into
$\mid\downarrow\rangle$. After pumping we take fast
'snapshots' of the EIT peak (50~sec $A^*$ scans,
$\Omega_{p}/2\pi= 25~{\rm MHz}$ and control at $A$ with
$\Omega_{c}/2\pi= 1~{\rm GHz}$). Between scans we kept the
system in the dark for 10 min. Figure~\ref{DNPpump} shows 6
subsequent snapshots. Right after pumping we observe a
blue-shifted and sharpened EIT peak ($T_2^*=3~{\rm ns}$).
This enhancement of $T_2^*$ probably results from
suppressed nuclear spin fluctuations, which generally
occurs when the polarization gets squeezed between a
polarizing and depolarizing mechanism with rates that are
both enhanced due to the DNP
 \cite{Xu2009,Latta2009a,vink2009natphys}.
The peak shift agrees in sign with Ref.~\cite{Kennedy2006}
but corresponds to $B_{nuc} = 21~{\rm mT}$ only (the ESR
studies \cite{Kennedy2006} and the work on dots easily
induced 200~mT - 1~T). Subsequent spectra show a clear
broadening of the EIT peak, which also shifts back to the
red. After about 1 hour, $T_2^*$ (Fig.~\ref{DNPpump},
inset) and the peak position stabilize at the values that
were observed before pumping. This agrees with the
relaxation time for DNP with $D^0$ systems
\cite{Kennedy2006}. Upon exploring how DNP occurs for
various EIT and pump conditions we found the effects to be
too weak for systematic control and drawing further
conclusions, and full understanding goes beyond the scope
of the present work. The work with dots showed that the
mechanism that dominates the DNP rate can be complex and
needs to account for driving-field assisted processes
\cite{Xu2009,Latta2009a}. We can nevertheless conclude that
our spin dephasing time is indeed limited by coupling to
nuclear spins.

%

In conclusion, we presented direct evidence that a $D^0$
ensemble in GaAs can be operated as a medium for EIT. The
electron spin dephasing time limits the quality of the EIT,
and is in the range $T_2^* \approx 2~{\rm ns}$ that results
from hyperfine coupling to fluctuating nuclear spins. The
EIT spectra form a sensitive probe for detecting how DNP
changes the fluctuations and the average of nuclear spin
polarization. However, direct optical driving of $D^0$
transitions yields much weaker DNP effects than in electron
spin resonance experiments with $D^0$ systems and related
EIT experiments on quantum dots, and a complete physical
picture of DNP effects in our system is not available.
Still, initial signatures of controlled DNP effects show
that the electron spin-dephasing time can be prolonged. Our
experimental approach is suited for exploring this further
in conjunction with experiments that aim to implement
various applications of EIT
\cite{Duan2001,Fleischhauer2005a}.

We thank B.~Wolfs, J.~Sloot and S.~Lloyd for contributions,
and the Dutch NWO and FOM, and the German programs
DFG-SPP~1285, BMBF~nanoQUIT and Research school of
Ruhr-Universit\"{a}t Bochum for support.

\end{document}